\documentclass{aastex61}

\accepted{February 8th, 2018}
\submitjournal{ApJ}

\shorttitle{The lithium depletion boundary in the Hyades.}
\shortauthors{Mart\'{\i}n et al.}

\begin{document}

\title{The lithium depletion boundary and the age of the Hyades cluster. \footnote{Based on data obtained at the 
Gran Telescopio Canarias}}

\correspondingauthor{Eduardo L. Mart\'{\i}n}
\email{ege@cab.inta-csic.es}

\author[0000-0002-1208-4833]{Eduardo L. Mart\'{\i}n}
\affil{Centro de Astrobiolog\'{\i}a (INTA-CSIC), Carretera de Ajalvir km 4,  E-28550 Torrej{\'o}n de Ardoz, Madrid, Spain}

\author{Nicolas Lodieu}
\affiliation{Instituto de Astrof\'{\i}sica de Canarias (IAC),  Calle V\'ia L\'actea s/n, E-38200 La Laguna, Tenerife, Spain}
\affiliation{Departamento de Astrof\'isica, Universidad de La Laguna (ULL), E-38205 La Laguna, Tenerife, Spain}
 
\author{Yakiv Pavlenko}
\affiliation{Main Astronomical Observatory of the National Academy of Sciences of Ukraine, Ukraine}

 \author{V\'{\i}ctor J. S. B\'ejar}
\affiliation{Instituto de Astrof\'{\i}sica de Canarias (IAC),  Calle V\'ia L\'actea s/n, E-38200 La Laguna, Tenerife, Spain}
\affiliation{Departamento de Astrof\'isica, Universidad de La Laguna (ULL), E-38205 La Laguna, Tenerife, Spain}



\begin{abstract}

Determination of the lithium depletion boundary (LDB), i.e., the observational limit below which the cores of very low-mass objects do not reach high enough temperature for Li destruction, has been used to obtain ages for several open clusters and stellar associations younger than 200 Myr, which until now has been considered as the practical upper limit on the range of applicability of this method. In this work we show that the LDB method can be extended to significant older ages than previously thought. Intermediate resolution optical spectra of six L-type candidate members in the Hyades cluster obtained using OSIRIS at the 10.4-m Gran Telescopio Canarias are presented. The Li~I 670.8~nm resonance doublet is clearly detected only in the two faintest and coolest of these objects, which are classified as L3.5 to L4 brown dwarf cluster members with luminosities around 10$^{-4}$ solar.  Lithium depletion factors are estimated for our targets with the aid of synthetic spectra and they are compared with predictions from evolutionary models. A LDB age of 650$\pm$70 Myr for the Hyades provides a consistent description of our data using  a set of state-of-the-art evolutionary models for brown dwarfs calculated by \citet{2015A&A..577..42}. 

\end{abstract}




\section{Introduction} \label{sec:intro}

The light element lithium (Li) is destroyed by collisions with protons in the cores of stars and high-mass brown dwarfs (BDs) 
with masses above 0.05 M$_{\odot}$ in timescales shorter than 1 Gyr \citep{1993ApJ...404L..17}. 
This nuclear process becomes observable as Li depletion in the surface via deep convective mixing. 
The discovery of Li in BD members in the Pleiades cluster \citep{1996ApJ...458..600, 1996ApJ...469L..53R}
was an observational confirmation that this light element is preserved in substellar mass objects as originally proposed by  \citet{1992ApJ...389L..83R}. The first attempt to detect Li in very low-mass members in open clusters (Hyades and Pleiades) was reported in 
\citet{1994ApJ...436..262}. Currently, the so-called Li test for BD candidates keeps on being used to constrain the ages and masses of young BDs in the field \citep{2017A&A...600..19}. 

The Li depletion boundary (LDB) in young open clusters is a chronometer that provides age estimates independent from 
other methods, such as turn off determination or main-sequence isochrone fitting. The physics involved in using the LDB are considered simpler than other methods used for dating clusters and associations \citep{2004ApJ...604..272B}. However, it has been considered that the LDB method is limited in its practical applicability to the age range between 20 Myr and 200 Myr, which implies that so far it could be used only in 2 stellar associations and 7 young open clusters \citep{1998ApJ...499..L199S, 2004ApJ...614..386, 2010ApJ...725L..111C, 2010MNRAS...409..1002, 2013MNRAS...434..2438}.  In this work we search for the LDB in the Hyades cluster, and we show that this method can be extended to significantly older ages than previously thought. 

Located at only 46.3$\pm$0.3 pc, the Hyades is the nearest open cluster \citep{2009A&A..497..209}. The isochrone age usually adopted for this cluster is 625$\pm$50 Myr \citep{1981A&A..93..136},  although a much older age of 1200 Myr  has been reported using evolutionary models with enhanced convective overshooting  \citep{1988A&A..193..148}. A recent estimate of the Hyades age using models that incorporate rotation has given 750$\pm$100 Myr \citep{2015ApJ..807..57}. Both convective overshooting and stellar rotation increase main-sequence lifetimes and thus yield older cluster ages. 
The Hyades cluster may be part of larger complex that also includes the Praesepe open cluster and a large number of field stars. This 
complex, known as Hyades supercluster, includes stellar populations with ages ranging from 0.5 Gyr to 1.0 Gyr 
\citep{1998AJ..116..284}.  

Even though the Hyades has a paucity of low-mass members with respect to younger clusters \citep{1992MNRAS..257..257}, which is understood as evidence for dynamical evolution that makes the lowest mass members disperse from the cluster core \citep{1987MNRAS..224..193}, 
a dozen L dwarf candidates with high probability of membership in the Hyades cluster were identifed from photometry and proper motion data presented in \citet{2008MNRAS...388..495}. Moreover, T dwarf candidate members were found by \citet{2008A&A..481..661}, and a late L dwarf member was recently reported in \citet{2017A&A..599..78}. 

Hyades brown dwarfs are benchmarks for understanding substellar evolution and they also provide a unique opportunity to extend the LDB method to a stellar cluster older than 200 Myr.  
In this paper, we present a search for the Li~I  resonance doublet centered at 670.8 nm in six  L dwarfs selected on the basis of confirmed L-type spectral classification from reconaissance spectra presented in \citet{2014MNRAS...445..3908}.

The rest of this paper is organized as follows: In Sect.~2 we present the target selection and the spectroscopic observations. 
In Sect.~3, we present the spectral types, spectrophotometric distances and radial velocities of our targets, and we estimate their membership probability in the Hyades. We also derive astrophysical parameters such as bolometric luminosities and effective temperatures using empirical calibrations. 
In Sect.~4, we present the search for Li, as well as for H$\alpha$ in emission, and we give the pseudo-equivalent width measurements of photospheric lines seen in the spectra. 
In Sect.~5, we discuss the determination of surface Li abundances in our sample of L dwarfs. Finally, in Sect.~6, 
we estimate the Hyades cluster age using the LDB method.

\section{Target selection and spectroscopic observations} \label{sec:obs}

We selected 6 targets from the list of 12 L-type Hyades brown dwarf candidates originally identified by \citet{2008MNRAS...388..495} on the basis of photometric and proper motion criteria. These objects had been confirmed as L-type dwarfs from low-resolution optical spectra \citep{2014MNRAS...445..3908}.  

All the spectra presented here were taken during dark nights in service mode at the 10.4-m Gran Telescopio Canarias (GTC) 
for programs GTC20-16B and GTC21-17B (PI, E. L. Mart\'{\i}n)
using the Optical System for Imaging and Low Resolution Integrated Spectroscopy (OSIRIS), \citet{2000SPIE...4008..623}. 
 The instrumental configuration was the long slit mode with R1000R grating, a slit width of 1.2 arcsec, and a detector bining of 2x2. 
 This combination gave a dispersion of 2.62~\AA / pixel, and a resolving power (R$=\lambda / \Delta \lambda$) of 561 at the central wavelength of 743 nm, corresponding to a nominal FWHM resolution of 12\AA . The actual FWHM resolution of the spectra ranged from 10\AA ~ to 12\AA ~according to measurements of sky airglow emission lines. The full spectral range of the data is from 510 nm to 1000 nm.  

The observing sequences consisted of one direct imaging exposure to acquire the target (30 s or 50 s exposure times, depending on sky brightness, through the Sloan $\it z$-band), 
another exposure through the same filter to check centering on the slit, and several exposures in the longslit spectroscopic mode. An offset of 15 arcsec along the slit direction was applied between each exposure.  All observations were carried out in parallactic angle. 

As a radial velocity (RV) reference the field L dwarf DENIS-P J0615493$-$010041 \citep{2008MNRAS...383..831} was observed on Dec 9th, 2016.  Since this object is brighter ($\it I$=17.0) than the Hyades targets, the exposure times were reduced to 20 s for the acquisition image and 900 s for each of the spectroscopic observations. Table~\ref{Obs} lists the targets observed for our lithium search and the observing details.  
 
We used IRAF to reduce the data. Bias subtraction and flat field correction were performed. One-dimensional spectra were extracted interactively using apsum. Wavelength calibration was made using an arc lamp spectrum obtained the same night as the science spectrum.  The spectro-photometric standard Hilt 600, observed with the same instrumental configuration in one of the same nights as the science spectra, was used to correct for instrumental response.

\startlongtable
\begin{deluxetable}{cccccc}
\tablecaption{Observing log \label{Obs}}
\tablehead{
\colhead{Abridged names}\tablenotemark{a} & \colhead{2MASS} &  \colhead{Date} & \colhead{Exposure time}\tablenotemark{b} 
& \colhead{Seeing}\tablenotemark{c} & \colhead{Weather conditions}\tablenotemark{d}  \\
\colhead{} & \colhead{coordinates} & \colhead{} & \colhead{seconds} & \colhead{arcsec} & \colhead{} \\}
\colnumbers
\startdata
Hya03           & J04102390+1459104 &  Feb 19th, 2017 &  2$\times$1700 & 1.0 &  Clear     \\
Hya08           & J04584566+1212343 &  Jan 27th, 2017  &  4$\times$1700 & 0.5 &  Clear      \\
Hya09           & J04463535+1451261 &  Dec 7th, 2016  &  2$\times$1500 & 0.6 &  Spectroscopic     \\
Hya09           & J04463535+1451261 &  Dec 25th, 2017  &  4$\times$2000 & 1.1 &  Clear     \\
Hya10           & J04173397+1430154 &  Jan 9th, 2018    &  2$\times$1800 & 1.1 &  Clear     \\
Hya11           & J03554191+2257016 &  Dec 7th, 2016   &  2$\times$1500 & 0.5 &  Spectroscopic     \\
Hya11           & J03554191+2257016  &  Jan 9th, 2018   &  2$\times$1800 & 1.1 &  Clear     \\
Hya12           & J04354302+1323448 &  Dec 24th, 2017  &  4$\times$2000 & 0.7 &  Clear     \\
DENIS-P J0615$-$01 & J06154934$-$0100415 &  Dec 9th, 2016  &  2$\times$900 & 0.8 &  Spectroscopic     \\
\enddata
\tablenotetext{a}{Names of Hyades L dwarfs from \citet{2008MNRAS...388..495}. }
\tablenotetext{b}{Number of exposures $\times$ on-target exposure time. }
\tablenotetext{c}{Values taken from GTC log of observations.}
\tablenotetext{d}{Sky conditions reported by the night observers. Spectroscopic means that some clouds were passing during the night.}
\end{deluxetable}

 \section{Cluster membership} \label{sec:mem}

In Figure 1 we show the calibrated spectra of our targets compared with spectral templates from the Sloan Digital Sky Survey (SDSS)  \citep{2010AJ...139..1808} that provided the best match to our data. 
 The spectral types adopted for our targets and their uncertainties are given in Table~\ref{tabSpT}. 
Our spectral types are systematically later by 0.5 to 2.0 subclasses than those published by \citet{2014MNRAS...445..3908}, and consequently the spectrophotometric distances of the objects are closer and more consistent with cluster membership. On the other hand for DENIS J0615-010 we find a spectral type of L1 which is 1.5 subclasses earlier than the L2.5 type reported by \citet{2008MNRAS...383..831}. 
Our spectral types ought to be more reliable than those estimated by other authors because of the higher quality of our data.

Spectrophometric distances were derived using the infrared photometry provided by \citet{2008MNRAS...388..495} and \citet{2014MNRAS...445..3908}, and the spectral type versus absolute magnitudes for field L dwarfs in \citet{2015ApJ...810..158}. They are given in Table~\ref{tabSpT}. We note that they are consistent with the range of distances expected for cluster members, i.e., between 32 pc and 60 pc according to \citet{2008MNRAS...388..495}. 
   
RV measurements were obtained for all the targets from cross correlation with the spectrum of the template DENIS J0615-010 using the IRAF task {\it fxcor}. 
Heliocentric radial velocity corrections were derived with the IRAF task {\it rvcor}. Instrumental zeropoint correction was applied using a radial velocity of $-$21.0 km/s for DENIS J0615$-$010 \citep{2015ApJS...220..18B}. 
The measured radial velocities obtained for our targets and their uncertainties are provided in Table~\ref{tabSpT}. 
The radial velocity values of all our Hyades targets are consistent within the error bars with the Hipparcos mean value for the cluster (39.5$\pm$0.3 km/s) as provided by \citet{2001A&A..367..111}. 

The contamination rate of the original sample of 12 Hyades L dwarf candidates discovered by \citet{2008MNRAS...388..495}  was estimated by those authors using comparison fields. They found that 2 candidates could be field M dwarfs, which was confirmed by \citet{2014MNRAS...445..3908} who found that Hya05 and Hya07 are indeed field M dwarfs. Hya02 was reported not to be a brown dwarf 
by \citet{2014MNRAS...441..2644}, but \citet{2014MNRAS...445..3908} did confirm it as an L dwarf. 
 
In order to estimate the membership probability of our targets in the cluster we assume that contamination rate of 2 objects has a poissonian error of $\pm$1.4 objects.  We assign a membership probability of 86\% to each of the 10 Hyades L dwarf candidates found by \citet{2008MNRAS...388..495} and confirmed as ultracool dwarfs by \citet{2014MNRAS...445..3908}, just based on the uncertainty in the 
contamination rate. To this probability we add the fact that each one of our targets have a RV value and an uncertainty, and we estimate 
the probability that this value is just a chance coincidence with the cluster RV value. To do this estimate we use the distribution of RV measurements for L dwarfs provided by \citet{2015ApJS...220..18B}. We find that among 21 L dwarfs with spectral subclass ranging from L1 to L5 one third of them have RV values in the range 20 to 60 km/s, which could be deemed as compatible with the cluster mean RV given the uncertainties of our RV measurements. Consequently, we increase the membership probability of our 4 targets from 86\% to 95\% because they have RV values consistent with the cluster within the measurement errors.

\startlongtable
\begin{deluxetable}{ccccccc}
\tablecaption{Hyades sample properties \label{tabSpT}}
\tablehead{
\colhead{Name}  &  \colhead{$SpT$} & \colhead{D$_{spec}$}\tablenotemark{a} 
& \colhead{$log$(L$_{\rm bol}$/L$_{\odot}$)}\tablenotemark{b} & \colhead{T$_{eff}$}\tablenotemark{b}  & \colhead{RV} & \colhead{Memb.} \\
\colhead{} & \colhead{} & \colhead{pc} & \colhead{} & \colhead{(K)} & \colhead{km/s} & \colhead{} \\}
\colnumbers
\startdata
Hya03           &   L2.0$\pm$0.5 &  48$\pm$9 & -3.83$\pm$0.13 &  1959 $\pm$113  & 24$\pm$13  & 95\%   \\
Hya08           &   L2.0$\pm$0.5 &  45$\pm$9 & -3.83$\pm$0.13 &  1959 $\pm$113  & 23$\pm$14   & 95\% \\
Hya09           &   L3.5$\pm$0.5 &  57$\pm$8 & -4.03$\pm$0.13 &  1758 $\pm$113  & 45$\pm$13   &  98\% \\
Hya10           &   L3.5$\pm$0.5 &  58$\pm$8 & -4.03$\pm$0.13 &  1758 $\pm$113  & 48$\pm$13   &  95\% \\
Hya11           &   L3.0$\pm$0.5 &  50$\pm$7 & -3.96$\pm$0.13 &  1822 $\pm$113  & 45$\pm$11   &  95\%  \\
Hya12           &   L4.0$\pm$0.5 &  56$\pm$8 & -4.10$\pm$0.13 &  1695 $\pm$113  & 36$\pm$13   &  98\% \\
\enddata
\tablenotetext{a}{Spectrophotometric distances derived from photometry in \citet{2008MNRAS...388..495} and 
\citet{2014MNRAS...445..3908} and the spectral type versus absolute magnitude relations for field L dwarfs \citep{2015ApJ...810..158}. }
\tablenotetext{b}{Bolometric luminosities and effective temperatures estimated using the calibrations of those parameters with spectral type given in \citet{2015ApJ...810..158}.}
\end{deluxetable}

All brown dwarf evolutionary models predict that Li should be preserved at masses below 0.06 M$_{\odot}$,  which corresponds to 
luminosities and temperatures in the realm of L dwarfs for ages younger than about 1 Gyr \citep{1984ApJS...491..856B, 1997ApJ...491..856B}. The presence of Li is therefore an indication of younger age than the average for field L dwarfs.  In the case of Hya09 and Hya12 we consider the presence of a detectable Li feature as an additional confirmation of membership in a cluster younger than the average age of the field population. Taking into account that the rate of Li detection in field L4 dwarfs is 40\%$\pm$11\% \citep{2008ApJ..689..1295}, we add this factor to their membership probability. Thus, the Hyades membership probabilities of the 6 targets considered in this study include the facts that they have the correct direction and magnitude of proper motion, the right spectral types and spectro-photometric distances, the correct location in color-magnitude diagrams, the correct radial velocity, and the presence of Li as an indicator of moderate youth and bona fide substellar status in the case of Hya09 and Hya12.  They are listed in Table~\ref{tabSpT}. 

\vspace{3mm}

\subsection{Astrophysical parameters} \label{sec:Par}

None of our targets show spectroscopic signs of being very young (age$<$100 Myr) as defined in \citet{2008ApJ..689..1295}. Their membership in the Hyades implies that they should have radii and surface gravities close to those of field L dwarfs. Therefore, it is justified to use the calibrations for field L dwarfs of known distance \citep{2015ApJ...810..158} to calculate their bolometric luminosities ($log$(L$_{\rm bol}$/L$_{\odot}$) and effective temperatures (T$_{eff}$) from the spectral types obtained by us. They are given in Table~\ref{tabSpT}. 

We also tried other approaches, such as using the observed photometry and bolometric corrections from \citet{2015ApJ...810..158}, 
and also Spectral Energy Distribution (SED) fitting with theoretical spectra following the procedures available in  VOSA \citep{2008A&A..492..277}, an interactive sofware developed by the Spanish Virtual Observatory.  These other approaches gave larger uncertainties and they could be affected by unresolved binarity of the targets, which is known to be at least 10\% among brown dwarfs in the Pleiades clusters \citep{2003ApJ...594..525, 2006ApJ...614..386}.   
 
The advantage of using directly the calibrations between spectral type and luminosity and temperature is that the results do not depend neither on the distance nor on the appartent brightness of the objects. Thus, we adopt those values for the analysis of the LDB in the Hyades.

\vspace{5mm}

\section{Atomic lines} \label{sec:Li}

We searched for the Li~I resonance doublet at 670.8 nm in all of the targets and did find something in 3 latest objects, namely Hya09, Hya11 and Hya12, as shown in Figure 2. The significance of these detections were estimated using the equation from \citet{1988IAUS...132..345}; namely,  $rms(pEW) = 1.6 \times (FWHM \times Disp.)^{1/2} / SNR $  where in our case the following parameters apply: FWHM=12 \AA , Disp.=2.62 \AA /pix, and the SNR per pixel for each spectrum was computed using the key m in the IRAF task splot across the wavelength range from 675 to 680 nm. These SNR estimates were more conservative than those found using other methods.   

In the case of Hya09 we found a predicted rms(pEW) for the Li~I resonance doublet at 670.78~nm~of 0.71 \AA , which formally implies a significance of 6.2$\pm$0.7$\sigma$ for the detection of this feature with a pEW=4.4$\pm$0.5 \AA  . 
Hya10 has the pooerest SNR of our sample and the predicted rms(pEW) for the Li~I resonance doublet from the Cayrel equation is 1.45 \AA . We do find a feature with a pEW of 3.2$\pm$0.8  \AA ~ at the position of the Li~I resonance doublet, but the significance is just 2.2$\pm$0.6$\sigma$, and hence we cannot claim to have a definitive Li detection in Hya10.  On the other hand, Hya12 has a Li~I feature with a pEW=8.5$\pm$0.4 \AA , which implies 
a confidence of 13.3$\pm$0.7$\sigma$ for this detection given than the predicted rms(pEW) is just 0.64 \AA , and it is indeed the clearest Li detection in our whole sample. For the rest of this paper we will consider as definitive Li detections only the features observed in Hya09 and Hya12, while the rest will be treated as upper limits. As a consistency check we also note that the FWHM of the Li~I feature in Hya09 and Hya12 were consistent with the expected FWHM resolution as measured in other photospheric lines. 

Independent measurements of the pEW for the Li~I features by all the co-authors of this work using the task splot confirmed that the uncertainties estimated from the Cayrel formula are realistic, although the interactive measurements tended to provide slightly higher 
error bars. Thus, we have adopted the more conservative uncertainties provided by the independent measurements and they are given in Table~\ref{tabEW}. Table~\ref{tabEW} summarizes the pEW measurements, upper limits and uncertainties. 

We also searched for H$\alpha$ in emission, an indicator of chromospheric activity but we could not find any clear detection among the Hyades L dwarfs. Upper limits to the H$\alpha$ pEW are given in Table~\ref{tabEW}. 
None of our Hyades targets have H$\alpha$ emission with pEW larger than about 4 \AA , which is consistent with recent estimates of about 10\%  H$\alpha$ emission detection rate among field L dwarfs \citep{2016ApJ...826..73}.  

Prominent photospheric lines in L dwarfs of Cs~I, Na~I and Rb~I were present in the observed spectral range. Their pEWs were measured in a manner analog to that described for the Li~I resonance feature. The line wavelengths listed in the NIST Atomic Spectra Database \citep{Kramida15} and the pEWs that we measured are given in Table~\ref{tabEW}.  For most of the lines there is good agreement between our 
pEWs and those listed in \citet{2015ApJS...220..18B} for DENIS J0615$-$010, with the notable exception of the Na~I doublet for which the pEWs in the literature are much weaker. This is likely due to the much higher spectral resolution (R$\sim$4100) of the data reported in \citet{2015ApJS...220..18B}, and highlights the importance of using data of similar resolution when comparing pEW values.  We also note that the Cs~I resonance line at 852.11 nm is stronger in Hya12 than in the other objects. This line is known to be very sensitive to T$_{eff}$ in L dwarfs, becoming stronger for cooler and later objects \citep{1999ApJ..519..802, 1999AJ...118..2466, 2000ApJ...538..363}, which is consistent with Hya12 being the coolest L dwarf in our sample. 

\startlongtable
\begin{deluxetable}{c|ccccccc}
\tablecaption{Pseudo equivalent width measurements\tablenotemark{a} \label{tabEW}}
\tablehead{
\colhead{Name} & \colhead{SNR} & \colhead{pEW H$\alpha$} & \colhead{pEW Li~I} & \colhead{pEW Rb~I} & \colhead{pEW Rb~I} &  \colhead{pEW Na~I} & \colhead{pEW Cs~I}  \\
\colhead{} & \colhead{675-680 nm} & \colhead{656.28 nm} & \colhead{670.78 nm} & \colhead{780.03 nm} & \colhead{794.76 nm} &  \colhead{818.32, 819.48 nm} &   \colhead{852.11 nm} \\
}
\colnumbers
\startdata
Hya03                    & 10.3  & $>$-3.5  & $<$0.5           & 2.5$\pm$0.3 & 1.6$\pm$0.3  & 5.7$\pm$0.3 & 1.4$\pm$0.2  \\
Hya08                    & 22.6  & $>$-0.8     & $<$0.3           & 2.8$\pm$0.1 & 2.6$\pm$0.2 & 4.8$\pm$0.4  & 1.6$\pm$0.1  \\
Hya09                    & 12.6  & $>$-1.2  & 4.4$\pm$0.5  & 2.7$\pm$0.2 & 4.3$\pm$0.3 & 4.3$\pm$0.4  & 2.7$\pm$0.1  \\
Hya10                    & 6.2    & $>$-2.0 & 3.2$\pm$0.8  & 2.6$\pm$0.3 & 4.2$\pm$0.4 & 5.6$\pm$0.6  & 2.0$\pm$0.1  \\
Hya11                    & 10.6  & $>$-2.7 & $<$0.8          & 2.4$\pm$0.2 & 3.7$\pm$0.2 & 4.1$\pm$0.3  & 2.2$\pm$0.1   \\
Hya12                    & 13.9  & $>$-2.8 & 8.5$\pm$0.4 & 8.0$\pm$1.0 & 6.0$\pm$0.7 & 5.0$\pm$0.3  & 3.7$\pm$0.1   \\
D J0615-01           & 28.5  & -2.3$\pm$0.2  & $<$0.1         &  2.7$\pm$0.1 & 2.6$\pm$0.2 & 6.7$\pm$0.3  & 1.6$\pm$0.2   \\
\enddata
\tablenotetext{a}{pEW measurements are given in \AA  .}
\end{deluxetable}

\vspace{2mm}

\section{Lithium abundances in Hyades L dwarfs} \label{sec:ab}

Previous studies of the LDB in young open clusters and associations did not attempt to estimate surface Li abundances because the gap in luminosity across the LDB was comparable to observational uncertainties. However, for increasing age the gap in luminosity from Li bearing BDs to Li naked BDs becomes larger (Figure 3) and this motivated us to estimate the Li abundances in our targets. We calculated theoretical pEWs 
for the Li~I resonance doublet using synthetic spectra for a range of T$_{eff}$, log g and log N(Li) values that are relevant for our targets. 
The model names used in Table~\ref{tabEW2} denote the T$_{eff}$ in K and log g adopted. 

The theoretical spectra were synthesized following the procedures described in \citet{2015A&A...581..73}. Fluxes across the Li~I resonance doublet are governed mainly by absorption in the extended wings of the K~I and Na~I resonance lines \citep{2000A&A..355..245}.  The TiO bands are of marginal strength in L dwarfs. In our computations only depletion of atomic lithium into molecular species (LiOH, LiCl, LiF) was accounted for. Depletion of lithium atoms into dust particles may reduce the number of absorbing particles in  the atmosphere. Unfortunately chemistry and kinematic properties of phase transition gas-dust are too poorly known to account for this effect.

\startlongtable
\begin{deluxetable}{ccc}
\tablecaption{Predicted Li abundance versus relative pEW Li~I \label{tabEW2}}
\tablehead{
\colhead{Model} & \colhead{$\Delta$log N(Li)} & \colhead{ (pEW Li~I)$_{depl}$/(pEW Li~I)$_{max}$}  \\ }
\colnumbers
\startdata
1600/5.0                    & $-$0.5    &  0.57     \\ 
1600/5.0                    & $-$1.0    &  0.32     \\ 
1600/5.0                    & $-$1.5    &  0.18     \\ 
1600/5.0                    & $-$2.0    &  0.10     \\ 
1600/5.0                    & $-$2.5    &  0.06     \\
1600/5.0                    & $-$3.0    &  0.04     \\
1600/4.5                    & $-$0.5    &  0.57     \\ 
1600/4.5                    & $-$1.0    &  0.33     \\
1600/4.5                    & $-$2.0    &  0.11     \\ 
1600/4.5                    & $-$3.0    &  0.04     \\
1800/5.0                    & $-$0.5    &  0.57     \\ 
1800/5.0                    & $-$1.0    &  0.33     \\ 
1800/5.0                    & $-$1.5    &  0.19     \\ 
1800/5.0                    & $-$2.0    &  0.11     \\ 
1800/5.0                    & $-$2.5    &  0.07     \\
1800/5.0                    & $-$3.0    &  0.04     \\ 
1800/4.5                    & $-$0.5    &  0.58     \\
1800/4.5                    & $-$1.0    &  0.33     \\ 
1800/4.5                    & $-$1.5    &  0.19     \\ 
1800/4.5                    & $-$2.0    &  0.11     \\ 
1800/4.5                    & $-$2.5    &  0.07     \\ 
1800/4.5                    & $-$3.0    &  0.04     \\
1800/4.5                    & $-$3.5    &  0.03      \\  
1800/4.5                    & $-$4.0    &  0.02     \\ 
2000/5.0                    & $-$0.5    &  0.57     \\
2000/5.0                    & $-$1.5    &  0.19     \\ 
2000/5.0                    & $-$2.0    &  0.11     \\ 
2200/5.0                    & $-$0.5    &  0.58     \\ 
2200/5.0                    & $-$1.5    &  0.19     \\ 
2200/5.0                    & $-$2.0    &  0.11     \\ 
\enddata
\end{deluxetable}

Hyades brown dwarfs presumably start off their evolution with an initial Li abundance of log N(Li) $\approx$3.3 (in the scale log N(H) = 12.0) 
which is common among newly born stars \citep{1960ApJ...131..83, 1991A&A...252..625} and A- and F-type cluster members \citep{2017AJ...153..128}. A high Li abundance of log N(Li) $=$3.25$\pm$0.25 has been inferred from self-consistent analysis of spectra of the field M9.5 brown dwarf LP944-20 \citep{2007MNRAS..380..1285} which has an age estimated to lie in the range 475--650 Myr \citep{1998MNRAS...296..L42}. 

Assuming an initial Li abundance of log N(Li) $=$3.3, we find that the predicted Li pEW values from the synthetic spectra are larger by a factor of 1.7$\pm$0.3 than the upper envelope of pEWs of the Li~I feature reported among L1.5--L4.5 dwarfs which is located between 8 and 15 \AA~  \citep{2000AJ..120..447, 2009AJ...137..3345, 2015A&A...581..73}.  It is likely that the synthetic spectra overpredict the strength of the Li~I feature because they do not include the effects of Li depletion into dust grains. In order to circumvent this problem we adopt the upper envelope of observed Li~I as representative of the initial Li abundance in L dwarfs and we use the synthetic spectra to calculate the relation between Li depletion due to Li burning and the ratio of the pEW of the Li~I feature with respect to the maximum value which corresponds to undepleted Li abundance. These relations are provided in Table~\ref{tabEW2}, and we note that they are rather insensitive to the T$_{eff}$ and log g adopted in the range of values considered in this work. 

From the pEW values given in Table~\ref{tabEW}, and the calculations presented in Table~\ref{tabEW2}, we derive the surface lithium abundances for our targets that are given in Table~\ref{tabAbun} assuming as undepleted Li abundance of log N(Li)$=$3.3. 
In particular, for an L4 dwarf (the spectral type of Hya12) we assume that the initial Li abundance corresponds to pEW (Li~I)$_{obs}=$11$\pm$4\AA~, and thus the pEW (Li~I) measured in Hya12 is a factor of 1.3$\pm$0.3 lower, which corresponds to a Li depletion of $\Delta$log N(Li)=$-$0.3$\pm$0.3 using the relations given in Table~\ref{tabEW2}. Using log N(Li)$=$3.3 as the initial Li abundance for the Hyades, we obtain a surface Li abundance of log N(Li)$=$3.0$\pm$0.3 for Hya12, which could be consistent with complete Li preservation in this object. On the other hand, for an L3.5 dwarf (the spectral type of Hya09) we assume that the initial Li abundance corresponds to pEW (Li~I)$_{obs}=$7$\pm$4\AA~, and thus the pEW (Li~I) measured in Hya09 is a factor of 1.6$\pm$0.6 lower, which corresponds to a Li depletion of $\Delta$log N(Li)=$-$0.5$\pm$0.5 using the relations given in Table~\ref{tabEW2}. Using log N(Li)$=$3.3 as the initial Li abundance for the Hyades, we obtain a surface Li abundance of log N(Li)$=$2.8$\pm$0.5 for Hya09, which is also consistent with no depletion at all, 
although with higher uncertainty than in the case of Hya12. 

\startlongtable
\begin{deluxetable}{cccc}
\tablecaption{Li depletion in Hyades L dwarfs \label{tabAbun}}
\tablehead{
\colhead{Name} & \colhead{$SpT$} &  \colhead{ (pEW Li~I)$_{obs}$/(pEW Li~I)$_{max}$}  & \colhead{log N(Li)}  \\ }
\colnumbers
\startdata
Hya03       &     L2.0$\pm$0.5  & $<$0.06              &  $<$0.8     \\ 
Hya08       &     L2.0$\pm$0.5  & $<$0.04              &  $<$0.3     \\ 
Hya09       &     L3.5$\pm$0.5  & 0.6$\pm$0.4       &  2.8$\pm$0.5 \\ 
Hya10       &     L3.5$\pm$0.5  & $<$1                   &   $<$3.3         \\ 
Hya11       &     L3.0$\pm$0.5  & $<$0.11              &  $<$1.3          \\
Hya12       &     L4.0$\pm$0.5  & 0.8$\pm$0.2       &  3.0$\pm$0.3 \\ 
\enddata
\end{deluxetable}

\vspace{6mm}

\section{The Hyades age} \label{sec:ab}

Given the bolometric luminosity, the effective temperature and the lithium abundance of an L dwarf, it should be possible to determine its age and mass, or place lower limits on them, using evolutionary models.   Our goal in this section is to combine the Li abundances and upper limits derived in our sample of L dwarfs to check whether or not we can find consistent age estimates for all of them using two independent sets of models, namely those from \citet{1997ApJ...491..856B} (hereafter Bu97) and those by \citet{2015A&A..577..42} (hereafter BHAC15).  

The predicted relations between Li depletion and luminosity or temperature for these two models are shown in Figures 3 and 4, respectively.  
The age limits obtained for the 5 Hyades L dwarfs from comparison of their luminosities, T$_{eff}$ and Li abundances with the two sets of models are given in Table~\ref{tabAge}. Hya10 is not included because it does not provide any additional constraint. 

Using the luminosities and the Bu97 models, we get a cluster age of 800$\pm$50 Myr. Using the temperatures, the same models yield 
975$\pm$25 Myr. On the other hand, the BHAC15 models give 635$\pm$85 Myr with the luminosities and 690$\pm$110 Myr with the temperatures. Thus, there is a significant region of agreement between the ages obtained from the BHAC15 luminosity tracks and the BHAC15 cooling curves, which is not the case for the Bu97 models.  This lack of self-consistency in the Bu97 models could be due to the fact 
that they  were mainly aimed at modeling in detail the properties of brown dwarfs cooler than 1300 K, whereas, for hotter brown dwarfs, such as those considered in this work, they use grey atmosphere approximations. The BHAC15 models, on the other hand, use detailed model atmospheres in the range of T$_{eff}$ values considered in this work, and we adopt them for estimating the Hyades LDB age.  

Assuming that the Hyades cluster is a coeval population of stars and brown dwarfs, it is found 
that an age of 650$\pm$70 Myr provides a consistent description of our results using the BHAC15 models.  
This result is in excellent agreement with the canonical Hyades age of 625$\pm$50 Myr \citep{1981A&A..93..136}, although it does not rule out a slightly older age, which may be consistent withing the error bars with the age of 750$\pm$100 Myr suggested by models that incorporate the effects of  stellar rotation \citep{2015ApJ..807..57}. 

The study of the LDB in the Hyades has some advantages and some disadvantages with respect to analog studies in younger clusters such as the Pleiades. On the positive side the LDB in the Hyades is less prone to magnetic activity effects because L dwarfs are less active than late-M dwarfs, and thus no activity corrections have been attempted in this work for the Hyades, whereas in younger clusters they have been deemed to be necessary \citep{2014ApJ...795..143J, 2015ApJ...813..108D}. Another bonus is that beyond 500 Myr the location of the LDB depends primarily on the cluster age and almost nothing on the mass of the brown dwarfs. On the other hand, the effects of dust condensation in L-type atmospheres and how they may impact on the observational properties of L dwarfs and the determination of the LDB are complicated and it could be worthwhile to be investigate them  in more detail. 


\startlongtable
\begin{deluxetable}{cccccccc}
\tablecaption{LDB age constraints for Hyades L dwarfs \label{tabAge}}
\tablehead{
\colhead{Name} & \colhead{$log$(L$_{\rm bol}$/L$_{\odot}$)} & \colhead{T$_{eff}$}  & \colhead{Li depletion} &  Age1 & Age2 & Age3 & Age4 \\ 
\colhead{} & \colhead{} & \colhead{K} & \colhead{} & \colhead{Myr} & \colhead{Myr} & \colhead{Myr} & \colhead{Myr} \\
\colhead{} & \colhead{} & \colhead{} & \colhead{} & \colhead{Bu97-L} & \colhead{Bu97-T} & \colhead{Ba15-L} & \colhead{Ba15-T} \\
}
\colnumbers
\startdata
Hya03       &    -3.83$\pm$0.13 &  1959 $\pm$113  & $>$99.0\%                &   $>$700    & $>$900      & $>$440 & $>$520    \\ 
Hya08       &    -3.83$\pm$0.13 &  1959 $\pm$113  & $>$99.9\%                &   $>$750  & $>$950    & $>$470 & $>$550    \\ 
Hya09       &    -4.03$\pm$0.13 &  1758 $\pm$113  & 0\%--90\%               & 450-940   &  550-1150  & 480--720 & 500--800    \\
Hya11      &     -3.96$\pm$0.13 &  1822 $\pm$113  & $>$95.0\%               &   $>$720    & $>$820      & $>$460 & $>$580    \\ 
Hya12       &    -4.10$\pm$0.13 &  1695 $\pm$113  & 0\%--78\%               & 500-850  &  600-1000 & 550-750 & 500--800    \\
\enddata
\end{deluxetable}

\vspace{3mm}

\acknowledgments

We thank the staff of the GTC who carried out the observations with OSIRIS in service mode for programs GTC20-16B and GTC21-17B, and I. Baraffe for providing a finer sampling of the Li depletion models for ages between 500 Myr and 700 Myr. ELM and NL are supported by grants 
AyA2015-69350-C3-1-P and AyA2015-69350-C3-2-P from the Spanish Ministry of Economy and Competitiveness 
(MINECO/FEDER). This publication makes use of VOSA, developed under the Spanish Virtual Observatory project supported from the Spanish MICINN through grant AyA2011-24052.  We thank the anonymous referee for numerous comments that helped to improve the manuscript.

%

\vspace{5mm}
\facilities{GTC(OSIRIS)}


\software{
          IRAF \citep{1986SPIE...627..733}
          VOSA \citep{2008A&A..492..277}, 
}



 \begin{figure}
\plotone{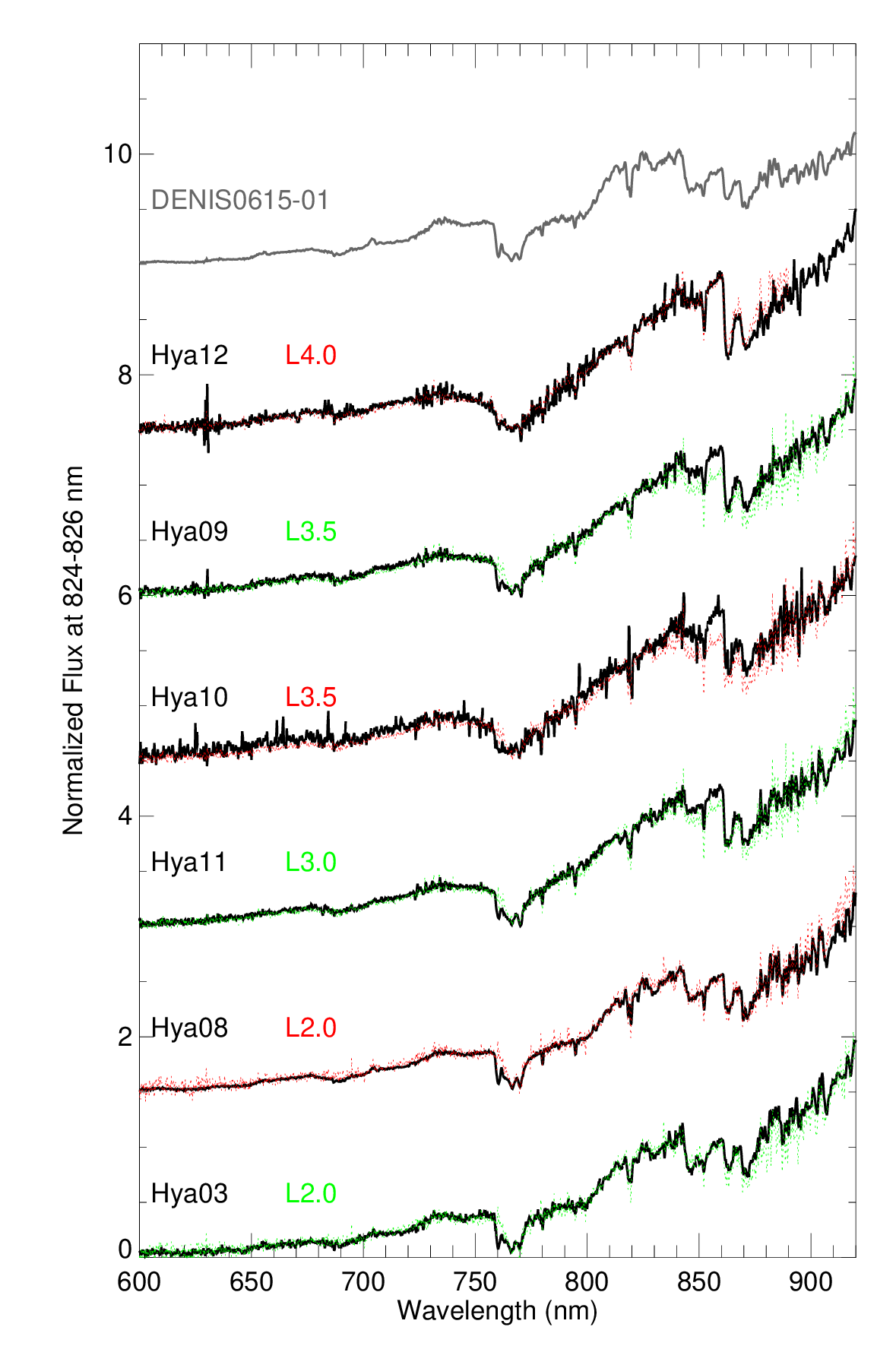}
\caption{\bf GTC/OSIRIS spectra of the Hyades L dwarfs observed in this work and the radial velocity standard. The best matching spectrum from the SDSS database is overplotted on each target and the spectral type of the template is labeled. 
\label{fig:fig1}}
\end{figure}

\begin{figure*}
\centering
\includegraphics[width=0.5\textwidth]{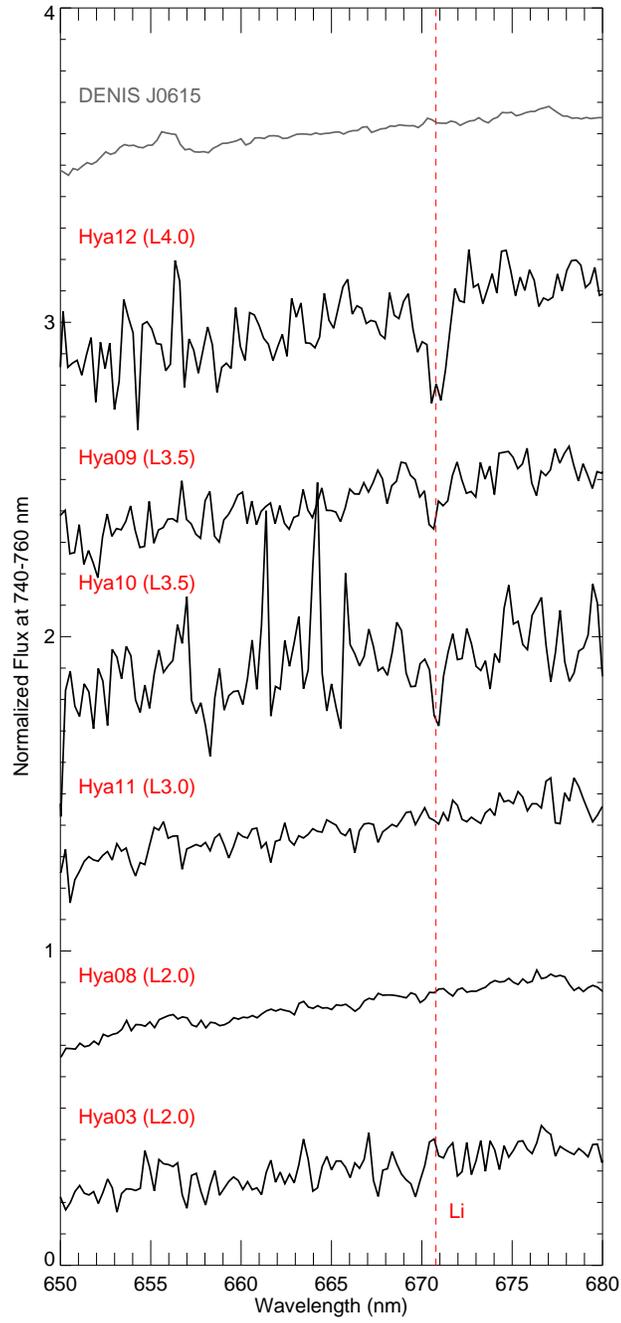}
\caption{\bf Zoom of the spectral region around the Li~I resonance doublet at 670.8~nm. The central wavelength of the Li feature is marked with a dashed vertical line.   
\label{fig:fig2}}
\end{figure*}

\begin{figure*}
\gridline{\fig{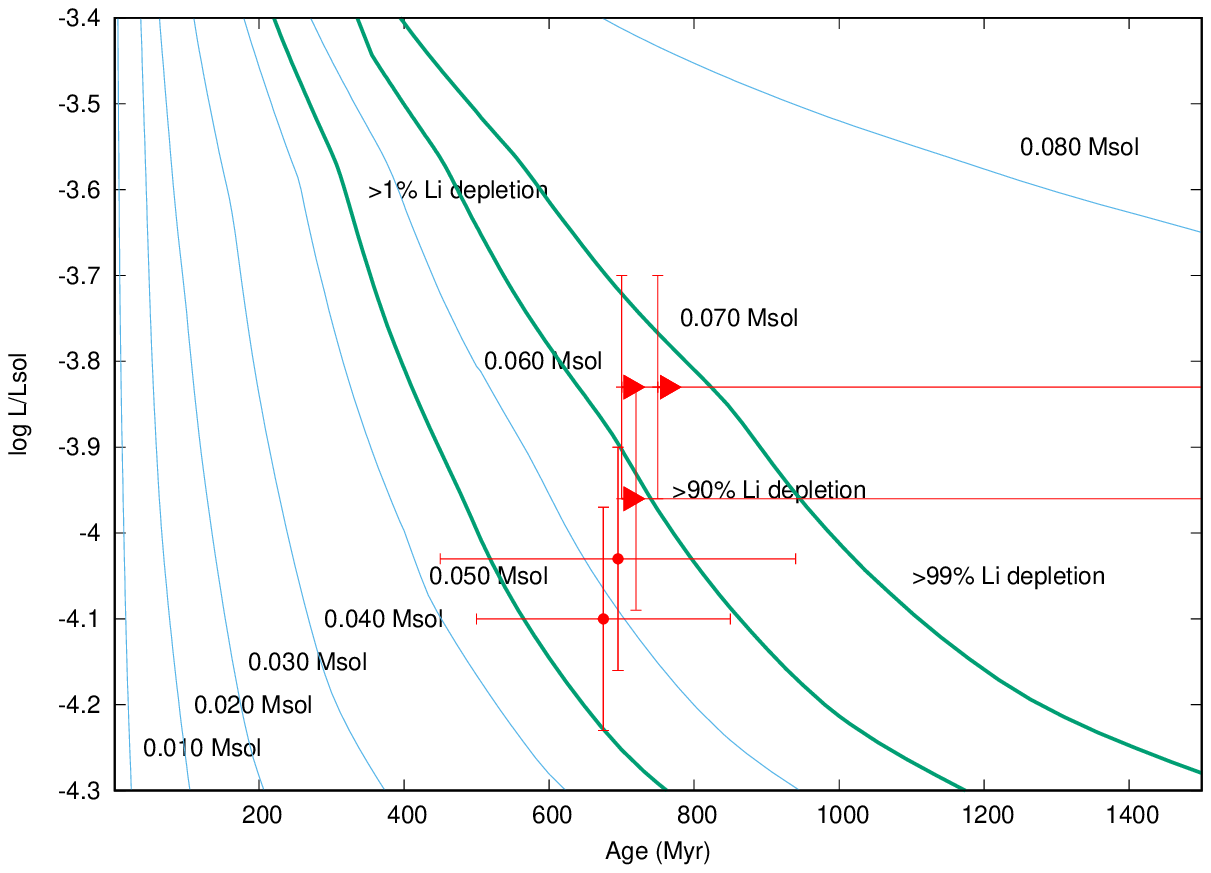}{0.5\textwidth}{(a)}
          \fig{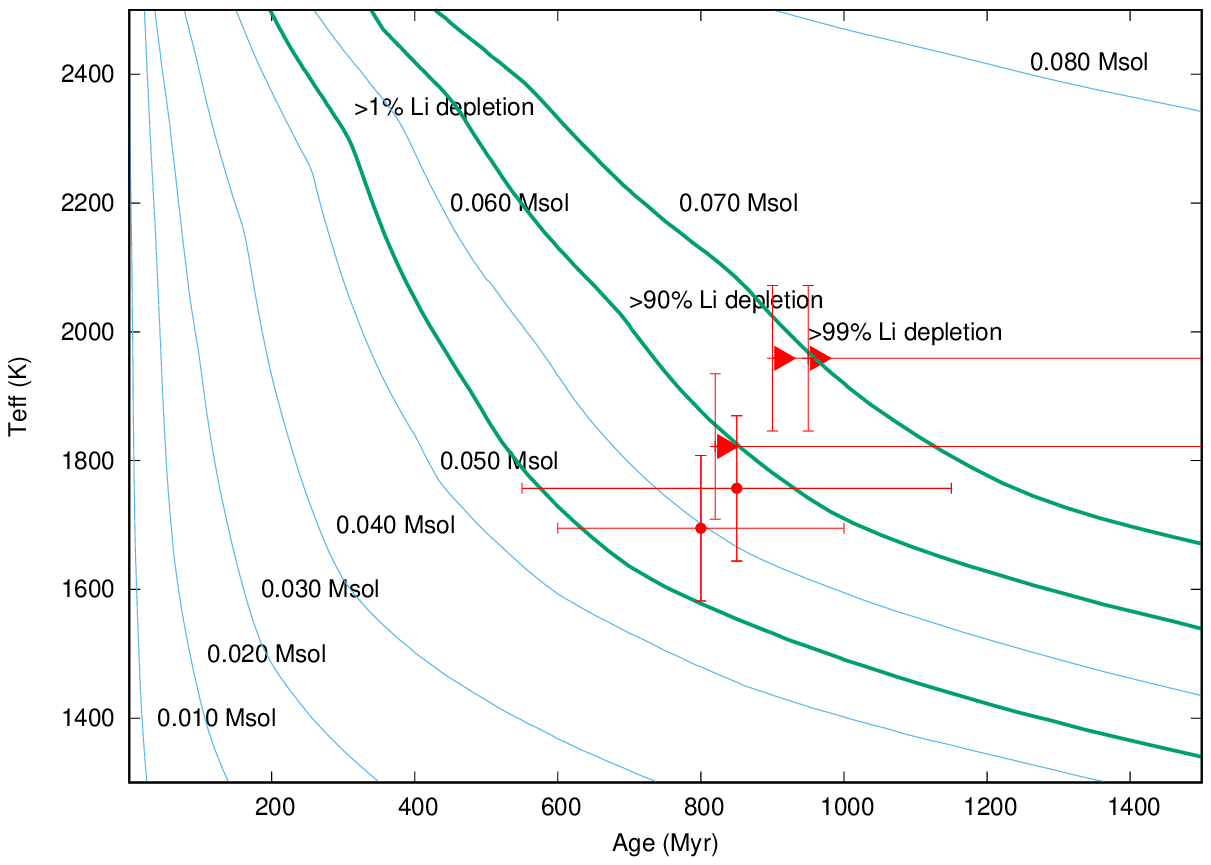}{0.5\textwidth}{(b)}
          }
\caption{\bf Evolutionary tracks for very low-mass stars and brown dwarfs as a function of age computed by \citet{1997ApJ...491..856B}.    
Masses and predicted Li depletion factors are labelled. The location of the Hyades L dwarfs observed in this work are shown.  
\label{fig:Bumodels}}
\end{figure*}

\begin{figure*}
\gridline{\fig{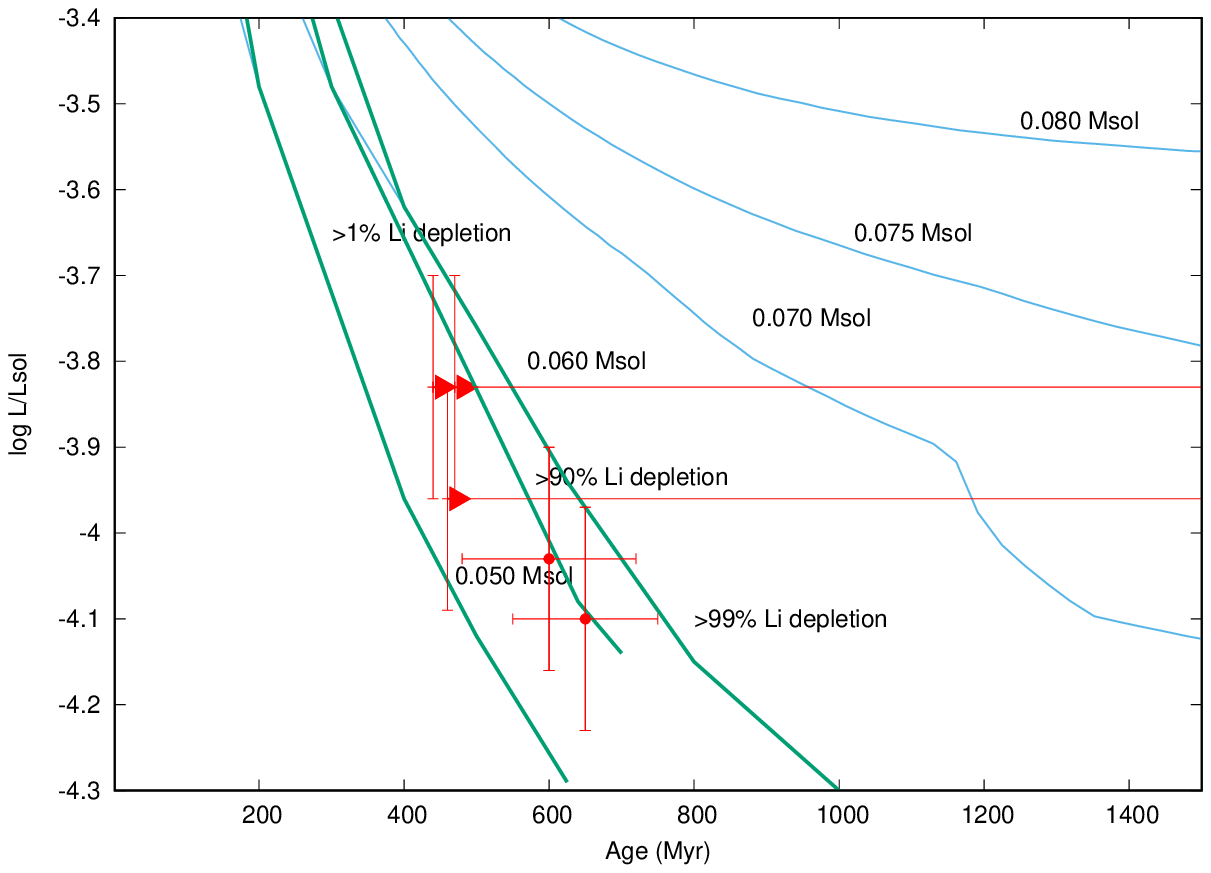}{0.5\textwidth}{(a)}
          \fig{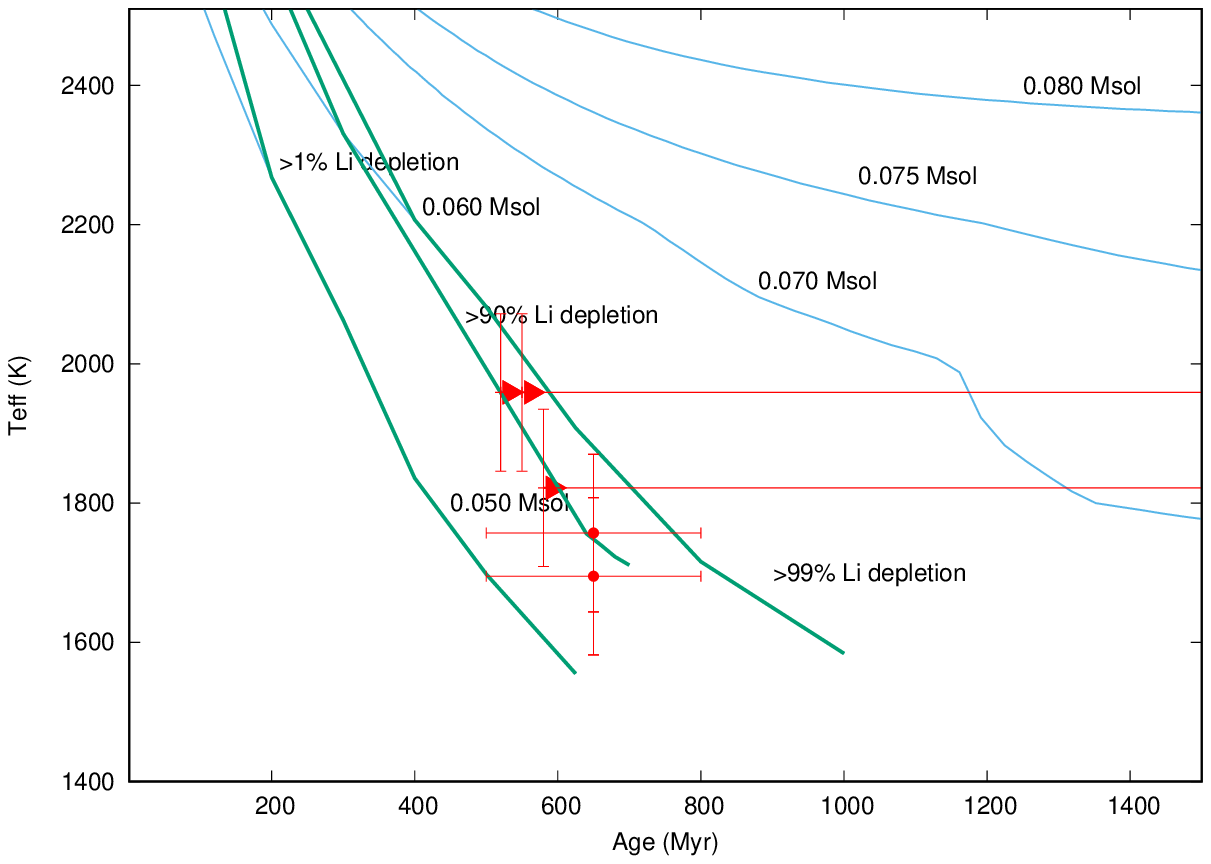}{0.5\textwidth}{(b)}
          }
\caption{\bf Evolutionary tracks for very low-mass stars and brown dwarfs as a function of age computed by \citet{2015A&A..577..42}.    
Masses and predicted Li depletion factors are labelled. The location of the Hyades L dwarfs observed in this work are shown.  
 \label{fig:Bamodels}}
\end{figure*}

\end{document}